\documentclass[aps,prd,reprint,floatfix,superscriptaddress,nofootinbib,longbibliography]{revtex4-2}
\usepackage{amsmath,amssymb,bm,graphicx,booktabs,multirow,siunitx,microtype,xcolor,hyperref}
\hypersetup{colorlinks=true,linkcolor=blue,citecolor=pink,urlcolor=pink}
\usepackage{url}
\usepackage{fontawesome5}
\usepackage{scalerel}
\usepackage{tikz}
\usetikzlibrary{svg.path} 
\definecolor{orcidlogocol}{HTML}{A6CE39}
\tikzset{
  orcidlogo/.pic={
    \fill[orcidlogocol] svg{M256,128c0,70.7-57.3,128-128,128C57.3,256,0,198.7,0,128C0,57.3,57.3,0,128,0C198.7,0,256,57.3,256,128z};
    \fill[white] svg{M86.3,186.2H70.9V79.1h15.4v48.4V186.2z}
                 svg{M108.9,79.1h41.6c39.6,0,57,28.3,57,53.6c0,27.5-21.5,53.6-56.8,53.6h-41.8V79.1z M124.3,172.4h24.5c34.9,0,42.9-26.5,42.9-39.7c0-21.5-13.7-39.7-43.7-39.7h-23.7V172.4z}
                 svg{M88.7,56.8c0,5.5-4.5,10.1-10.1,10.1c-5.6,0-10.1-4.6-10.1-10.1c0-5.6,4.5-10.1,10.1-10.1C84.2,46.7,88.7,51.3,88.7,56.8z};
  }
}

\newcommand\orcidicon[1]{\href{https://orcid.org/#1}{\mbox{\scalerel*{
\begin{tikzpicture}[yscale=-1,transform shape]
\pic{orcidlogo};
\end{tikzpicture}
}{|}}}}

\definecolor{mycolor}{RGB}{0,0,204}
\definecolor{pink}{RGB}{255,0,127}

\usepackage{tabularx}
\usepackage{graphicx}
\newcommand{\mpl}{M_{\rm Pl}}
\newcommand{\cO}{\mathcal O}
\newcommand{\chis}{\chi_t}
\newcommand{\weff}{\bar w_{\rm re}}

\newcommand{\ord}[1]{\mathcal O\!\left(#1\right)}
\begin{document}

\title{Dilaton-Flattened Axion Inflation }

 \author{Pirzada \orcidicon{0009-0002-2274-9218}}
 
 \email{pirzada@itp.ac.cn}%
\affiliation{CAS Key Laboratory of Theoretical Physics, Institute of Theoretical Physics, Chinese Academy of Sciences, Beijing 100190, China}
\affiliation{School of Physical Sciences, University of Chinese Academy of Sciences, No. 19A Yuquan Road, Beijing 100049, China}
\author{Ali Muhammad}
\email{alimuhammad@phys.qau.edu.pk}
\affiliation{CAS Key Laboratory of Theoretical Physics, Institute of Theoretical Physics, Chinese Academy of Sciences, Beijing 100190, China}
\affiliation{School of Physical Sciences, University of Chinese Academy of Sciences, No. 19A Yuquan Road, Beijing 100049, China}

  \author{Tianjun Li }%
 \email{tli@itp.ac.cn}%
 \affiliation{School of Physics, Henan Normal University, Xinxiang 453007, P. R. China   }%
 \author{Imtiaz Khan}
\email{ikhanphys1993@gmail.com}
\affiliation{Department of Physics, Zhejiang Normal University, Jinhua, Zhejiang 321004, China}
\affiliation{Research Center of Astrophysics and Cosmology, Khazar University, Baku, AZ1096, 41 Mehseti Street, Azerbaijan}
\affiliation{Zhejiang Institute of Photoelectronics, Jinhua, Zhejiang 321004, China}
 \author{Mussawir Khan}%
 \email{mussawirkhan@ihep.ac.cn}
\affiliation{State Key Laboratory of Particle Astrophysics, Institute of High Energy Physics, Chinese Academy of Sciences, Beijing 100049, China}
\affiliation{University of Chinese Academy of Sciences, Beijing 100049, China}

\begin{abstract}
We present a solvable same-sector effective theory for anomaly-inspired axion inflation, in which a heavy trace-anomaly mode dynamically backreacts on the axion potential. The tree-level elimination of the radial field resums the backreaction into a closed-form Lambert-$W$ potential, naturally flattening the hilltop potential without external plateau operators. By deriving the exact trough metric, we evaluate all the observables on the fully reduced one-field action, bypassing uncontrolled kinetic approximations. Calibrated at $N_\star=56$, reheating-compatible branches yield $r\simeq0.033$--$0.036$ and $\alpha_s\simeq-(4.6$--$4.7)\times10^{-4}$, comfortably satisfying the current ACT/SPT/BICEP constraints. The evolution remains strictly adiabatic ($m_\perp^2/H^2\gtrsim6.1$, $\Omega/H\lesssim7.6\times10^{-4}$) with negligible sound-speed and metric corrections. We provide  analytic control over the constant-$w_{\rm eff}$ reheating map, the $N_{\rm re}=0$ boundary, and robustness against vacuum-offset deformations. This Lambert-$W$ backbone establishes a precise, deformable benchmark for confining axion inflation, with microscopic matching and reheating microphysics accessible as systematic EFT refinements.
\end{abstract}

\maketitle

\section{Introduction}

Natural inflation remains theoretically attractive because a shift symmetry can protect the scalar potential over super-Hubble evolution, while confinement naturally supplies non-perturbative vacuum energy around the required magnitude. The original natural inflation construction and its modern descendants remain a standard benchmark for large-field model building~\cite{Freese:1990rb,Freese:2004un,Planck:2018jri,Stein:2021yma,Zhang:2018gni}. At the same time, current CMB data increasingly pressure the unflattened branch of ordinary natural inflation through the tensor amplitude unless the effective decay constant is large, and this has focused attention on mechanisms that reduce the local curvature of the potential without forfeiting calculability~\cite{Dong:2010in,Nomura:2017rik,Nomura:2017tm,Gatica:2026pni}. One particularly economical possibility is that the same strongly coupled sector, which generates the axion potential, also contains a heavier scalar degree of freedom whose backreaction flattens the effective potential. While recent statistical updates \cite{Gatica:2026pni}  confirm that large-N pure natural inflation can marginally survive the latest ACT bounds, those constructions rely on multi-branch vacuum structure to achieve flattening. In contrast, our same-sector Lambert-W mechanism generates identical phenomenological flattening through the dynamical backreaction of a heavy trace-anomaly mode inside a calculable local EFT, without invoking large-N branch structure. As in standard natural inflation, the present construction retains a super-Planckian effective axion decay constant; the ultimate resolution of any quantum-gravity field-range constraints is therefore delegated to the UV completion, as discussed in \ref{sec:discussion}.

A closely related issue arises in anomaly-inspired composite and glueball inflation. The trace anomaly of a confining gauge theory already furnishes a natural radial sector, and nonminimally coupled glueball/dilaton inflation has been analyzed from this perspective in earlier work~\cite{Bezrukov:2011rq,Channuie:2012kn,Khan:2026nsz,Anguelova:2016qju,pirzada:2026glt}. On the topological side, the large-$N$ structure of Yang--Mills vacuum energy and its effective-Lagrangian realizations have long suggested that the $\theta$ dependence and heavy glueball sector should not be thought as unrelated ingredients~\cite{Witten:1980sp,Veneziano:1979ec,Halperin:1998hq,Fugleberg:1998kk}. The present question is narrower and more concrete: if the trace-anomaly mode and topological vacuum energy are retained simultaneously inside one local effective theory, can the heavy radial response be eliminated in closed form, and does the resulting same-sector potential generate a phenomenologically viable flattened branch with explicit dynamical control?

A useful analytic answer emerges within a local heavy-dilaton effective theory. The precise statement proved below is not that the full Yang--Mills vacuum functional has been derived from first principles. Rather, once one adopts the same-sector local EFT specified in Sec.~\ref{sec:EFT}, the radial degree of freedom can be integrated out in closed form at tree level. The resulting axion potential is controlled by a Lambert-$W$ function~\cite{Corless:1996zz,Veberic:2012ax}; the perturbative series of heavy-field corrections is resummed into one analytic expression; the hilltop curvature, orthogonal-mode mass ratio, and the local turn-rate parameter along the trough can each be written explicitly. We also show that the same solution furnishes analytic control over the reduced sound-speed correction, over the induced trough metric familiar from curved-valley EFT reductions~\cite{Shiu:2011qw,Cespedes:2012hu,Burgess:2012dz}, and over the sensitivity to generic local EFT deformations. In what follows, we therefore reserve the strongest claims for this closed-form local EFT reduction rather than for a microscopic Yang--Mills derivation.

The current observational motivation is immediate. ACT DR6 reports $n_s=0.9666\pm0.0077$ for ACT alone and $n_s=0.9709\pm0.0038$ for the combined Planck+ACT likelihood, while the latest SPT-3G D1 analysis enters the same observational territory, with the CMB-SPA combination summarized by $n_s=0.9684\pm0.0030$~\cite{Louis:2025fju,Camphuis:2025spt}. The official BICEP/Keck bound remains $r_{0.05}<0.036$ at $95\%$ confidence~\cite{BICEPKeck:2024txt}. These data motivate a precise question: can a same-sector backreaction mechanism lower $r$ relative to ordinary natural inflation while preserving a controlled single-clock description? The analytic formulae derived below sharpen that question at the EFT level, even though they do not replace a full likelihood analysis.

The remainder of the paper is organized as follows. Section~\ref{sec:EFT} formulates the same-sector local EFT and derives the Lambert-$W$ potential. Section~\ref{sec:exact} derives the hilltop curvature, the orthogonal-mode mass ratio, the geometric turn-rate expression, and a perturbative control formula for generic local EFT deformations, retaining a nonzero vacuum offset analytically throughout. Section~\ref{sec:inflation} confronts the backreacted potential with current ACT/SPT/BICEP-Keck targets, presents benchmark trajectories, develops robustness bands around the observationally relevant windows, and quantifies the sensitivity to modest nonzero $B$. Section~\ref{sec:reheating} derives the reheating relations for the backreacted potential in the constant-$\weff$ approximation, identifies the $N_{\rm re}=0$ boundary, and gives the corresponding numerical consequences. Section~\ref{sec:discussion} discusses the relation of the present construction to pure natural inflation, heavy-field flattening, and the scope of the effective description.

\section{Same-sector local effective theory and closed-form backreaction resummation}
\label{sec:EFT}

We work with the minimal local two-field effective potential
\begin{equation}
\begin{aligned}
U(\sigma,\theta) &= V_0 + \frac12 m_\sigma^2\,\delta\sigma^2 + \chi_0 e^{-b\delta\sigma/\mpl} S(\theta),\\
S(\theta) &\equiv 1-\cos\theta,
\end{aligned}
\label{eq:Ustart}
\end{equation}
where $\delta\sigma\equiv\sigma-\sigma_0$ and $\theta\equiv a/f_a$. The parameter $\chi_0\equiv\chis(\sigma_0)$ sets the overall topological scale, and the dimensionless coefficient $b$ controls the response of the susceptibility to the heavy radial displacement. Equation~\eqref{eq:Ustart} should be regarded as a same-sector local EFT: the trace-anomaly motivation follows the classic effective description of gluodynamics~\cite{Migdal:1982jp,Novikov:1983jt}, while the topological component is patterned on the $\theta$-dependence of confining Yang--Mills vacuum energy and on effective-Lagrangian treatments in which heavy glueball variables can be integrated out to recover a multi-branch $\theta$ potential~\cite{Witten:1980sp,Veneziano:1979ec,Halperin:1998hq,Fugleberg:1998kk,Nomura:2017rik}. The exponential susceptibility is therefore an effective ansatz motivated by the scaling role of the dilaton, not a first-principles derivation of the microscopic Yang--Mills path integral. The virtue of Eq.~\eqref{eq:Ustart} is that it isolates the same-sector backreaction mechanism in a form that is sufficiently constrained to be solved in closed form, while still general enough to admit a systematic local-deformation analysis below.

It is convenient to introduce the dimensionless variables
\begin{equation}
x \equiv \frac{b\,\delta\sigma}{\mpl},
\qquad
\beta \equiv \frac{b^2\chi_0}{2 m_\sigma^2 \mpl^2},
\qquad
B \equiv \frac{2\beta V_0}{\chi_0}.
\label{eq:defs}
\end{equation}
In terms of $x$, the potential is
\begin{equation}
U(x,\theta)=\chi_0\left[\frac{B}{2\beta}+\frac{x^2}{4\beta}+e^{-x}S(\theta)\right].
\label{eq:Ux}
\end{equation}
The trough condition is algebraic,
\begin{equation}
\frac{\partial U}{\partial x}=\chi_0\left(\frac{x}{2\beta}-e^{-x}S\right)=0
\quad\Longrightarrow\quad
x e^x = 2\beta S(\theta).
\label{eq:stationarity}
\end{equation}
Therefore the radial displacement along the trough is solved by the principal Lambert branch,
\begin{equation}
x(\theta)=W\!\left(2\beta S(\theta)\right).
\label{eq:Wsol}
\end{equation}
Substituting Eq.~\eqref{eq:Wsol} into Eq.~\eqref{eq:Ux} yields the closed-form backreaction-resummed effective potential,
\begin{equation}
U_{\rm eff}(\theta)=V_0+\frac{\chi_0}{2\beta}\left[W+\frac12 W^2\right],
\qquad
W\equiv W\!\left(2\beta S(\theta)\right).
\label{eq:Ueff}
\end{equation}
Equation~\eqref{eq:Ueff} is a tree-level statement inside the local EFT; no slow-roll approximation enters its derivation. In this sense it provides a same-sector realization of the heavy-field flattening logic studied in other inflationary settings, but with the flattening encoded by a solvable radial response rather than by an externally prescribed higher-dimensional operator~\cite{Dong:2010in,McAllister:2008hb}. Expanding in small $\beta S$ gives
\begin{equation}
U_{\rm eff}(\theta)=V_0+\chi_0\Bigl[S-\beta S^2+2\beta^2S^3-\frac{16}{3}\beta^3S^4+\cO(\beta^4S^5)\Bigr]~.~
\label{eq:series}
\end{equation}
Thus, the negative quartic harmonic found in the perturbative analysis is simply the leading term of the resummation. The backreacted potential thus organizes the entire heavy-field expansion into a single analytic function.

\begin{figure}[t]
    \includegraphics[width=\linewidth]{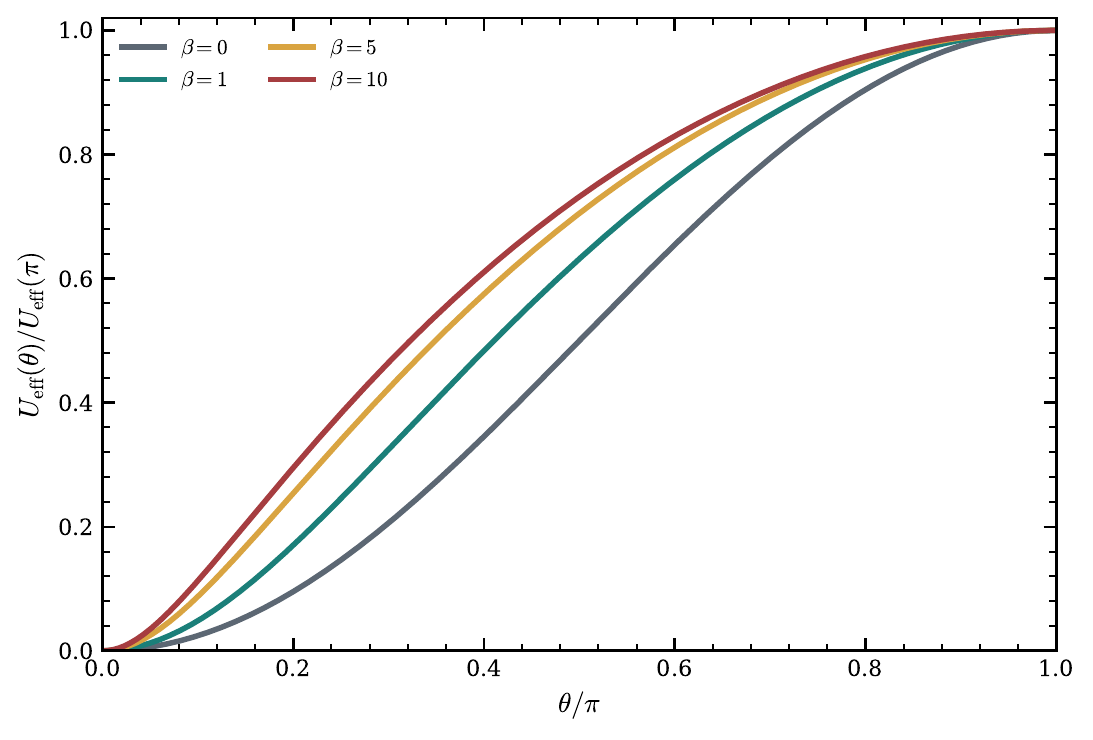}
    \caption{Normalized backreacted effective potential $U_{\rm eff}(\theta)/U_{\rm eff}(\pi)$ for $\beta=0,1,5,10$. The $\beta\to0$ branch reproduces ordinary natural inflation, while increasing $\beta$ implements same-sector flattening through the radial backreaction encoded in Eq.~\eqref{eq:Ueff}.}
    \label{fig:potential}
\end{figure}

Figure~\ref{fig:potential} shows the normalized shape of the backreacted potential for representative values of $\beta$. The flattening is monotonic: increasing $\beta$ reduces the hilltop curvature and broadens the quasi-plateau region while preserving the same periodic topology as ordinary natural inflation.

\section{Analytic local observables of the trough}
\label{sec:exact}

\subsection{Hilltop curvature with nonzero vacuum offset}

\begin{figure}[t]
    \includegraphics[width=\linewidth]{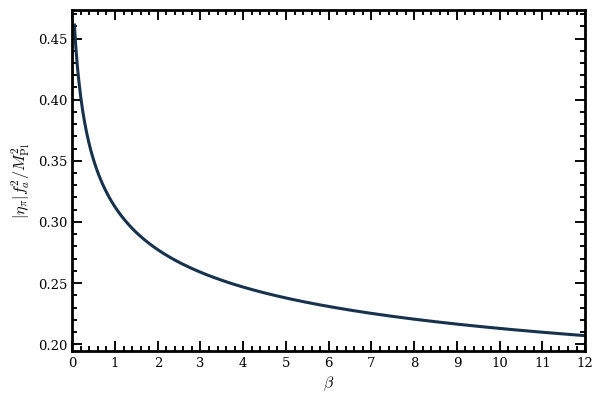}
    \caption{ hilltop-curvature suppression factor $|\eta_\pi|f_a^2/\mpl^2 = [2+W(4\beta)]^{-1}$ on the tuned branch $B=0$. The decrease is monotonic and controlled entirely by the same-sector backreaction parameter $\beta$.}
    \label{fig:hilltop}
\end{figure}

Because $U_{,\theta}(\pi)=0$, the leading local shape information at the hilltop is carried by the second derivative. From Eq.~\eqref{eq:Ueff},
\begin{equation}
U_{,\theta\theta}(\pi)=-\chi_0 e^{-W_\pi},
\qquad
W_\pi\equiv W(4\beta).
\label{eq:Udblhill}
\end{equation}
Using $e^{-W_\pi}=W_\pi/(4\beta)$ and Eq.~\eqref{eq:Ueff}, the hilltop curvature parameter can be written in slow-roll form as
\begin{equation}
\begin{aligned}
\eta_\pi \equiv \mpl^2\frac{U_{,aa}(\pi)}{U_{\rm eff}(\pi)}
&=\frac{U_{,\theta\theta}(\pi)}{f^2 U_{\rm eff}(\pi)} \\
&= -\frac{W_\pi}{2f^2\left(B+W_\pi+\frac12 W_\pi^2\right)},
\end{aligned}
\label{eq:etapiB}
\end{equation}
where $f\equiv f_a/\mpl$. For the tuned inflationary branch used in the numerical figures, $B=0$, and Eq.~\eqref{eq:etapiB} reduces to
\begin{equation}
\eta_\pi = -\frac{1}{f^2\left(2+W_\pi\right)}.
\label{eq:etapi}
\end{equation}
Hence the curvature-suppression factor is
\begin{equation}
\frac{|\eta_\pi|f_a^2}{|\eta_\pi|_{\beta=0}f_a^2}=\frac{1}{2+W(4\beta)}.
\label{eq:hilltopfactor}
\end{equation}
Equation~\eqref{eq:hilltopfactor} is closed-form within the local EFT and makes the flattening manifest. Figure~\ref{fig:hilltop} displays the suppression factor as a function of $\beta$.

\subsection{Orthogonal-mode mass ratio}

The orthogonal fluctuation is governed by the curvature of the two-field potential transverse to the trough. Since $x=b\delta\sigma/\mpl$, the canonical radial mass is obtained from the $x$-Hessian,
\begin{equation}
\frac{\partial^2 U}{\partial x^2}=\chi_0\left(\frac{1}{2\beta}+e^{-x}S\right).
\label{eq:Uxx}
\end{equation}
On the trough, Eq.~\eqref{eq:stationarity} implies $e^{-x}S=x/(2\beta)=W/(2\beta)$, so
\begin{equation}
\left.\frac{\partial^2 U}{\partial x^2}\right|_{\rm tr} = \chi_0\frac{1+W}{2\beta}.
\label{eq:Uxxtr}
\end{equation}
Because $\partial/\partial\sigma=(b/\mpl)\partial/\partial x$, the orthogonal mass is
\begin{equation}
m_\perp^2 = \left(\frac{b}{\mpl}\right)^2\left.\frac{\partial^2 U}{\partial x^2}\right|_{\rm tr}
= m_\sigma^2(1+W).
\label{eq:mperp}
\end{equation}
Using $3H^2=U_{\rm eff}/\mpl^2$, one obtains the local mass ratio
\begin{equation}
\frac{m_\perp^2}{H^2}=3b^2\frac{1+W}{B+W+\frac12 W^2}.
\label{eq:mratioB}
\end{equation}
For $B=0$ this reduces to
\begin{equation}
\frac{m_\perp^2}{H^2}=3b^2\frac{1+W}{W\left(1+\frac12 W\right)}.
\label{eq:mratio}
\end{equation}
Equation~\eqref{eq:mratioB} also makes the parametric role of $b$ completely explicit. If one demands a target adiabaticity floor
\begin{equation}
\frac{m_\perp^2}{H^2}\ge \mu,
\label{eq:mufloor}
\end{equation}
then the required radial-response coefficient is bounded by
\begin{equation}
b\ge \left[\frac{\mu\left(B+W+\frac12 W^2\right)}{3(1+W)}\right]^{1/2}.
\label{eq:bfloor}
\end{equation}
On the tuned branch $B=0$, the conservative requirement $\mu=5$ at the hilltop gives
\begin{equation}
b\ge1.55\;(\beta=5),\qquad b\ge1.69\;(\beta=10),
\label{eq:bfloorbench}
\end{equation}
so the benchmark choice $b=2$ is not a tuned threshold value but a modestly conservative point in parameter space.
At the hilltop, Eq.~\eqref{eq:mratio} gives the lower bounds quoted in the abstract: for $b=2$,
\begin{equation}
\begin{aligned}
\left.\frac{m_\perp^2}{H^2}\right|_{\pi}
&=\frac{12(1+W_\pi)}{W_\pi(1+W_\pi/2)} \\
&=8.30\;\; (\beta=5),\qquad 7.00\;\; (\beta=10).
\end{aligned}
\label{eq:hilltopmratio}
\end{equation}
Over the observable window the bounds are stronger, as shown below.

\subsection{Turn-rate control along the trough}

\begin{figure}[t]
    \includegraphics[width=\linewidth]{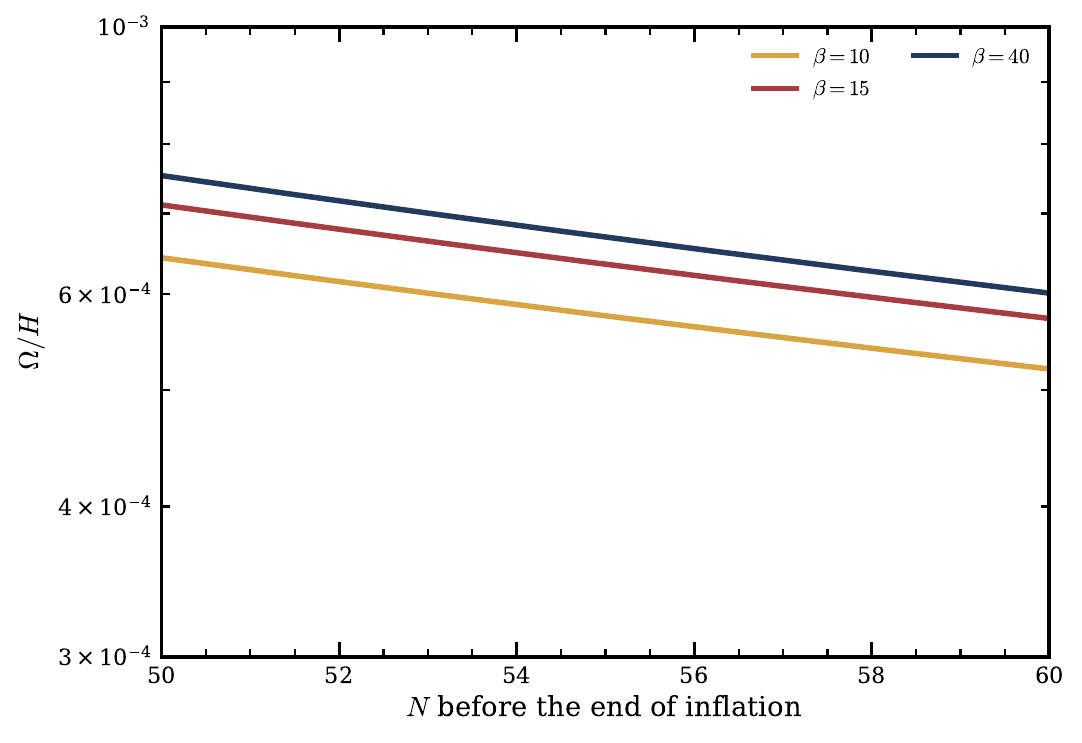}
    \caption{Turn-rate ratio $\Omega/H$ over the observable window $N=50$--$60$ for the same  trough-reduced benchmark branches. The maximal values are of order $10^{-4}$--$10^{-3}$, placing the displayed trajectories in a strongly adiabatic regime.}
    \label{fig:mass_turn}
\end{figure}

The same local EFT also allows the geometric bending of the trough to be characterized analytically. In dimensionless canonical variables
\begin{equation}
y\equiv \frac{\delta\sigma}{\mpl}=\frac{x}{b},
\qquad
z\equiv \frac{a}{\mpl}=f\,\theta,
\qquad f\equiv \frac{f_a}{\mpl},
\label{eq:canoncoords}
\end{equation}
Eq.~\eqref{eq:Wsol} implies
\begin{equation}
\frac{dy}{dz}=\frac{1}{bf}\frac{W}{1+W}\cot\frac{\theta}{2},
\label{eq:dz1}
\end{equation}
and, after one further differentiation,
\begin{equation}
\frac{d^2 y}{dz^2}=\frac{1}{bf^2}\frac{W\left(\cos\theta-W^2-2W\right)}{\left(1-\cos\theta\right)(1+W)^3}.
\label{eq:dz2}
\end{equation}
The curvature of the trough in field space is therefore
\begin{equation}
\kappa(\theta)=\frac{\left|d^2y/dz^2\right|}{\left[1+\left(dy/dz\right)^2\right]^{3/2}},
\label{eq:kappa}
\end{equation}
and in slow roll the dimensionless turn rate is
\begin{equation}
\frac{\Omega}{H}=\sqrt{2\epsilon}\,\kappa(\theta).
\label{eq:OmegaH}
\end{equation}
Equations~\eqref{eq:dz1}--\eqref{eq:OmegaH} provide a second analytic consistency check. Figure~\ref{fig:mass_turn} shows both $m_\perp^2/H^2$ over the full $0\le N\le 60$ interval and $\Omega/H$ over the observable window $50\le N\le 60$ for the reheating-compatible benchmark branches introduced below. For the  trough-reduced $\beta=10$, $15$, and $40$ solutions with $b=2$, one finds
\begin{equation}
\begin{aligned}
\left.\frac{m_\perp^2}{H^2}\right|_{N=50\text{--}60}
&\gtrsim 7.86\; (\beta=10), \\
&\gtrsim 7.08\; (\beta=15), \\
&\gtrsim 6.05\; (\beta=40),
\end{aligned}
\label{eq:mobs}
\end{equation}
while the turn rate satisfies
\begin{equation}
\begin{aligned}
\left.\frac{\Omega}{H}\right|_{N=50\text{--}60}
&\lesssim 6.43\times10^{-4}\; (\beta=10), \\
&\lesssim 7.12\times10^{-4}\; (\beta=15), \\
&\lesssim 7.53\times10^{-4}\; (\beta=40).
\end{aligned}
\label{eq:turnobs}
\end{equation}
These branches therefore lie in a strongly adiabatic regime over the observationally relevant part of inflation. This statement can be sharpened using the standard adiabatic EFT relation
\begin{equation}
\begin{aligned}
c_s^{-2}&\simeq 1+\frac{4\Omega^2}{M_{\rm eff}^2},\\
M_{\rm eff}^2&=m_\perp^2-\Omega^2+\ord{\epsilon H^2},
\end{aligned}
\label{eq:cs}
\end{equation}
valid for slowly varying turns with a parametrically heavy orthogonal mode~\cite{Achucarro:2012sm,Achucarro:2012yr,Cespedes:2012hu,Cespedes:2013rda}. In the displayed branches, $\epsilon$ stays at the usual slow-roll level while $m_\perp^2/H^2\gtrsim6$ and $(\Omega/H)^2\lesssim6\times10^{-7}$ over the observable window, so the omitted $\ord{\epsilon H^2}$ term remains parametrically and numerically subleading compared with $m_\perp^2$. Combining Eqs.~\eqref{eq:mobs} and \eqref{eq:turnobs} therefore gives the conservative bounds
\begin{equation}
\begin{aligned}
c_s^{-2}-1
&\lesssim 2.2\times10^{-7}\; (\beta=10), \\
&\lesssim 2.9\times10^{-7}\; (\beta=15), \\
&\lesssim 3.8\times10^{-7}\; (\beta=40),
\end{aligned}
\label{eq:csbounds}
\end{equation}
over the observable window. In the present benchmarks the reduced single-field description is therefore quantitatively indistinguishable from a unit-sound-speed trajectory at the level relevant for the background observables quoted below. This does not replace a full transfer-matrix calculation for a microscopic UV completion, but it does provide a local check that the displayed branches are well inside the adiabatic regime emphasized in Refs.~\cite{Achucarro:2010jv,Achucarro:2012sm,Achucarro:2012yr,Shiu:2011qw,Burgess:2012dz,Cespedes:2013rda}.

\subsection{Exact trough-reduced kinetic factor}

\begin{figure}[t]
    \includegraphics[width=\linewidth]{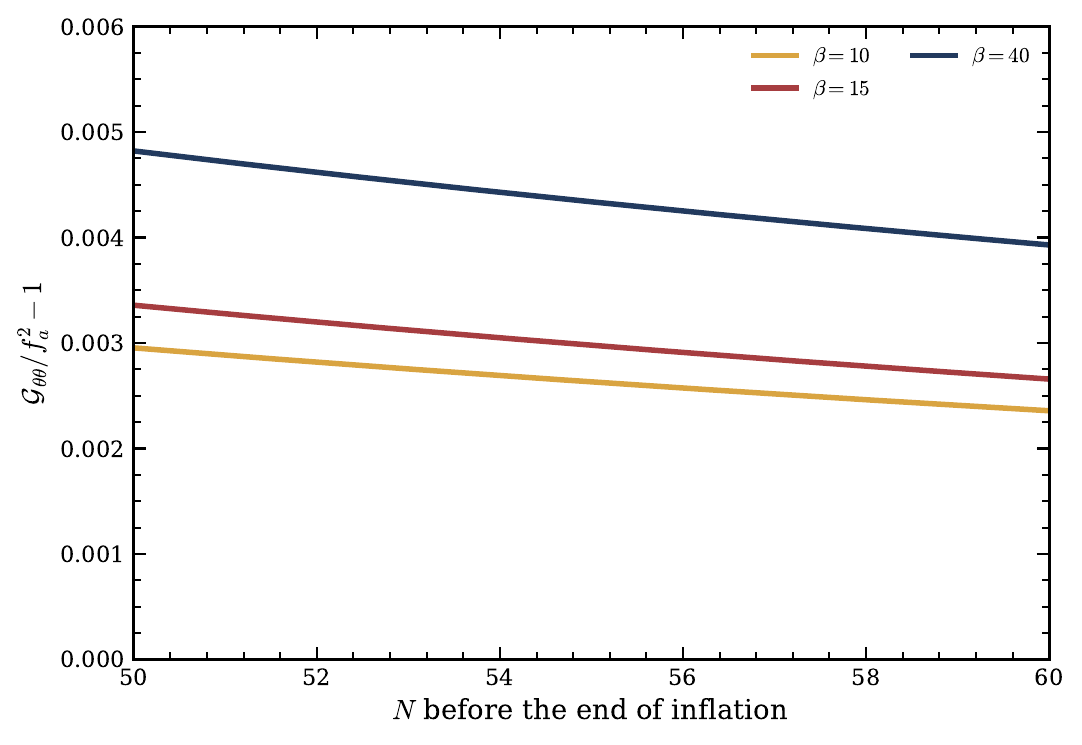}
    \caption{Exact trough-metric correction $\mathcal G_{\theta\theta}^{\rm tr}/f_a^2-1$ over the observable window for the reheating-compatible benchmark branches. The correction remains below the half-percent level, but it is retained exactly in the benchmark observables quoted below.}
    \label{fig:metriccorr}
\end{figure}

Integrating out the heavy mode at the background level also induces a nontrivial line element along the trough. Since $\sigma_{\rm tr}(\theta)=\sigma_0+(\mpl/b)W(2\beta S)$, one has
\begin{equation}
\left(\frac{d\sigma_{\rm tr}}{d\theta}\right)^2=\frac{\mpl^2}{b^2}\left(\frac{W}{1+W}\cot\frac{\theta}{2}\right)^2,
\label{eq:dsigdtheta}
\end{equation}
so the  one-field action obtained by restricting the two-field theory to the trough takes the form
\begin{equation}
\begin{aligned}
\mathcal L_{\rm tr}={}&\frac12\mpl^2\,\mathcal G_{\theta\theta}^{\rm tr}(\theta)(\partial\theta)^2-U_{\rm eff}(\theta),\\
\mathcal G_{\theta\theta}^{\rm tr}(\theta)={}&f^2+\frac{1}{b^2}\left(\frac{W}{1+W}\cot\frac{\theta}{2}\right)^2.
\end{aligned}
\label{eq:Gtr}
\end{equation}
This is the curved-valley reduction familiar from multi-field trough EFTs~\cite{Langlois:2008qf,Shiu:2011qw,Burgess:2012dz,Senatore:2010wk,Gong:2016uoa}. The canonically normalized trough field $\varphi_{\rm tr}$ is therefore defined by
\begin{equation}
\frac{d\varphi_{\rm tr}}{d\theta}=\mpl\sqrt{\mathcal G_{\theta\theta}^{\rm tr}(\theta)}.
\label{eq:canontr}
\end{equation}
The  trough-reduced slow-roll parameters used for the benchmark phenomenology below are then
\begin{equation}
\epsilon_{\rm tr}=\frac{1}{2\mathcal G_{\theta\theta}^{\rm tr}}\left(\frac{U_{,\theta}}{U}\right)^2,
\qquad
\eta_{\rm tr}=\frac{1}{U}\left[\frac{U_{,\theta\theta}}{\mathcal G_{\theta\theta}^{\rm tr}}-\frac{\left(\mathcal G_{\theta\theta}^{\rm tr}\right)_{,\theta}}{2\left(\mathcal G_{\theta\theta}^{\rm tr}\right)^2}U_{,\theta}\right],
\label{eq:srtr}
\end{equation}
with the usual first-order expressions $n_s\simeq1-6\epsilon_{\rm tr}+2\eta_{\rm tr}$ and $r\simeq16\epsilon_{\rm tr}$. Figure~\ref{fig:metriccorr} shows the size of the induced metric correction for the reheating-compatible benchmark branches adopted later. The correction is small, remaining below the half-percent level across the observable window, but it is straightforward to retain exactly and we do so for the benchmark observables, the adiabaticity diagnostics quoted in Eqs.~\eqref{eq:mobs}--\eqref{eq:csbounds}, and the reheating tables in Sec.~\ref{sec:reheating}. This  trough reduction removes any dependence of the headline benchmark numbers on the minimal constant-kinetic approximation.

\subsection{Controlled deformations of the local EFT}

The analytic formulas above rely on the minimal ansatz in Eq.~\eqref{eq:Ux}. It is therefore useful to state explicitly how a generic local deformation propagates into the trough solution. Write the dimensionless potential as
\begin{equation}
u(x,\theta)\equiv \frac{U(x,\theta)}{\chi_0}
= \frac{B}{2\beta}+\frac{x^2}{4\beta}+e^{-x}S(\theta)+\Delta u(x,\theta),
\label{eq:udef}
\end{equation}
where $\Delta u$ collects cubic and higher radial self-interactions, corrections to the exponential susceptibility, and any other local terms that are perturbative in the EFT sense; correspondingly, $\Delta u_{,x}$ denotes the derivative of the deformation with respect to the dimensionless radial variable $x$. Let $x_0(\theta)=W(2\beta S)$ denote the trough of the undeformed theory, so $u_{0,x}(x_0,\theta)=0$. Expanding the deformed stationarity condition around $x_0$ gives
\begin{equation}
\begin{aligned}
\delta x(\theta)\equiv x_{\rm tr}-x_0
&= -\frac{\Delta u_{,x}(x_0,\theta)}{u_{0,xx}(x_0,\theta)}+\ord{\Delta u^2} \\
&= -\frac{2\beta}{1+x_0}\,\Delta u_{,x}(x_0,\theta)+\ord{\Delta u^2},
\end{aligned}
\label{eq:deltax}
\end{equation}
where we used $u_{0,xx}(x_0,\theta)=(1+x_0)/(2\beta)$. The effective potential on the deformed trough is then
\begin{equation}
\begin{aligned}
u_{\rm eff}^{\rm def}(\theta)
&= u_{\rm eff}^{(0)}(\theta)+\Delta u(x_0,\theta) \\
&\quad -\frac{\beta}{1+x_0}\left[\Delta u_{,x}(x_0,\theta)\right]^2+\ord{\Delta u^3}.
\end{aligned}
\label{eq:ueffdef}
\end{equation}
Equations~\eqref{eq:deltax} and \eqref{eq:ueffdef} are useful because they separate two logically distinct issues. The Lambert-$W$ solution remains the closed-form zeroth-order backbone of the theory, while the local sensitivity to deviations from the minimal ansatz is controlled by the same denominator $1+x_0=1+W(2\beta S)$ that also governs the orthogonal stiffness. In practical terms, the benchmark predictions obtained from Eq.~\eqref{eq:Ueff} are stable whenever
\begin{equation}
\left|\frac{2\beta\,\Delta u_{,x}}{1+x_0}\right|\ll 1,
\qquad
\left|\Delta u(x_0,\theta)\right|\ll u_{\rm eff}^{(0)}(\theta),
\label{eq:defcond}
\end{equation}
through the CMB window. This provides a compact analytic criterion for judging when cubic/quartic radial corrections or modest departures from a pure exponential susceptibility modify the benchmark observables only perturbatively rather than qualitatively.

\begin{figure}[t]
    \includegraphics[width=\linewidth]{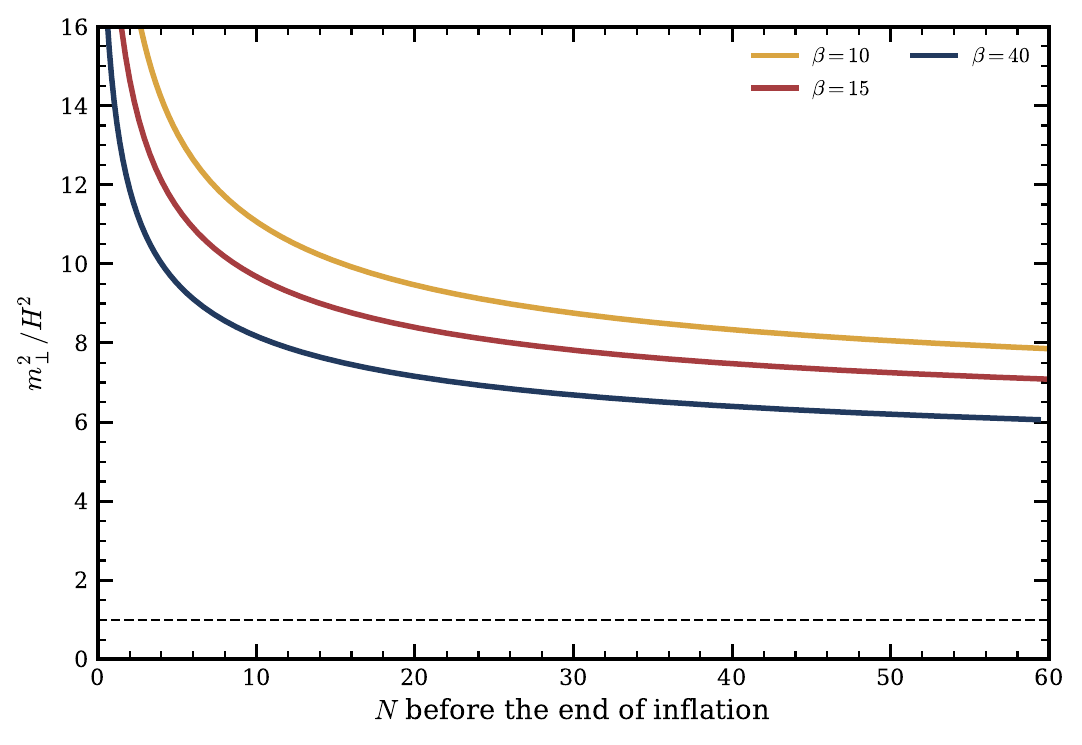}
    \caption{Orthogonal-mode mass ratio $m_\perp^2/H^2$ along the full background evolution for the reheating-compatible  trough-reduced benchmark branches with $b=2$. The ratio remains parametrically larger than unity throughout the displayed interval.}
    \label{fig:mratiofig}
\end{figure}

\section{Inflationary dynamics and current observational targets}
\label{sec:inflation}

\begin{figure}[t]
    \includegraphics[width=\linewidth]{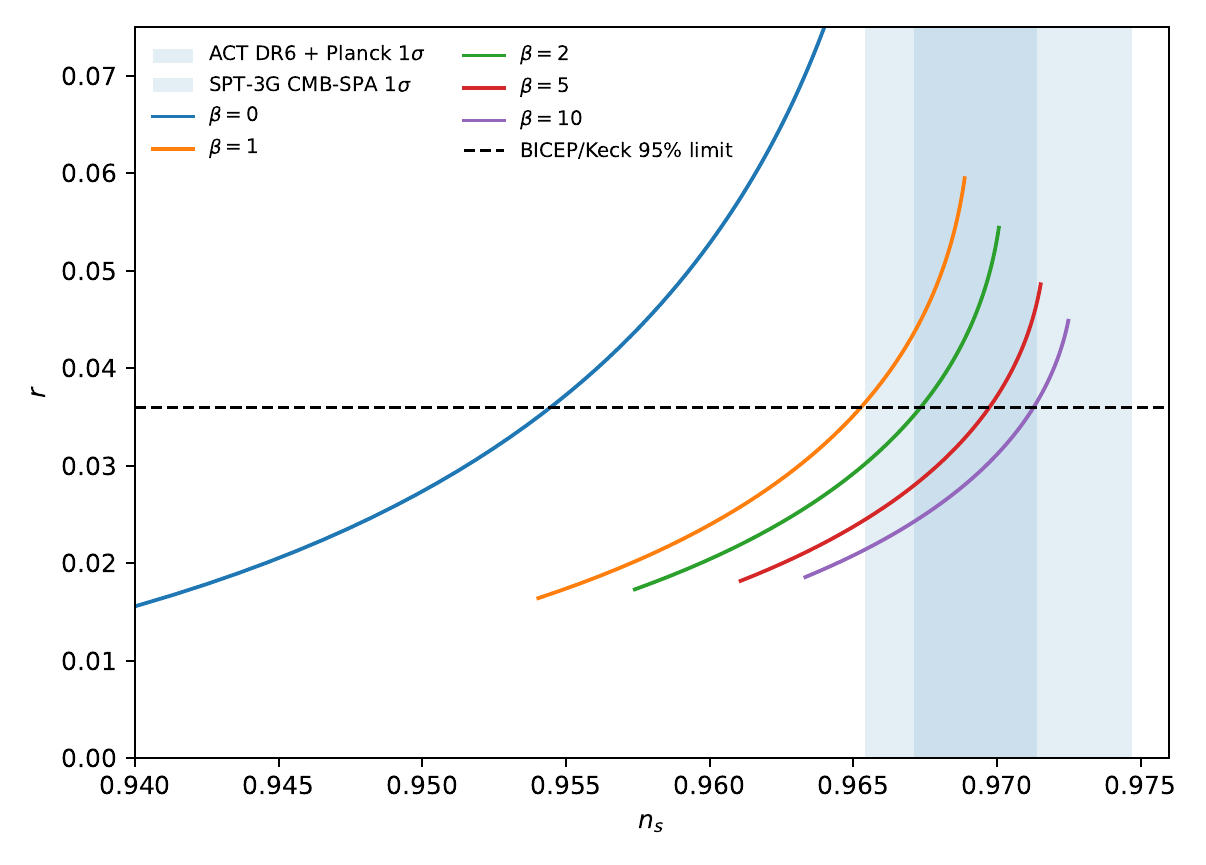}
    \caption{Model $n_s$--$r$ trajectories at fixed $N_\star=60$ for several values of the backreaction parameter $\beta$. The shaded vertical bands indicate the ACT+Planck and SPT-3G/CMB-SPA $1\sigma$ scalar-tilt intervals, and the horizontal dashed line shows the BICEP/Keck bound $r_{0.05}<0.036$. Increasing $\beta$ flattens the potential and lowers the tensor amplitude at fixed tilt.}
    \label{fig:nsr}
\end{figure}

\begin{figure}[t]
    \includegraphics[width=\linewidth]{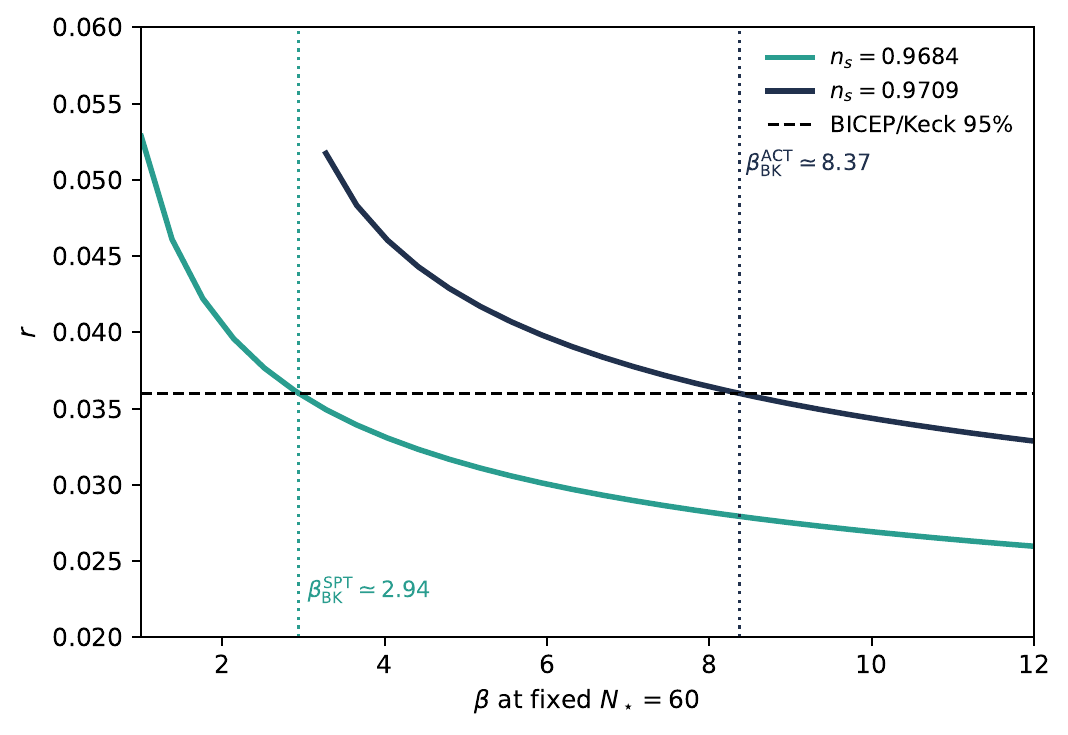}
    \caption{Tensor amplitude at fixed $N_\star=60$ and fixed scalar tilt as a function of the same-sector flattening parameter $\beta$. The intersections with the BICEP/Keck bound give the fixed-tilt calibration conditions quoted in Eq.~\eqref{eq:thresholds}. These curves are closed-form EFT predictions but are not a substitute for a full likelihood analysis.}
    \label{fig:thresholdfig}
\end{figure}

\begin{figure}[t]
    \includegraphics[width=\linewidth]{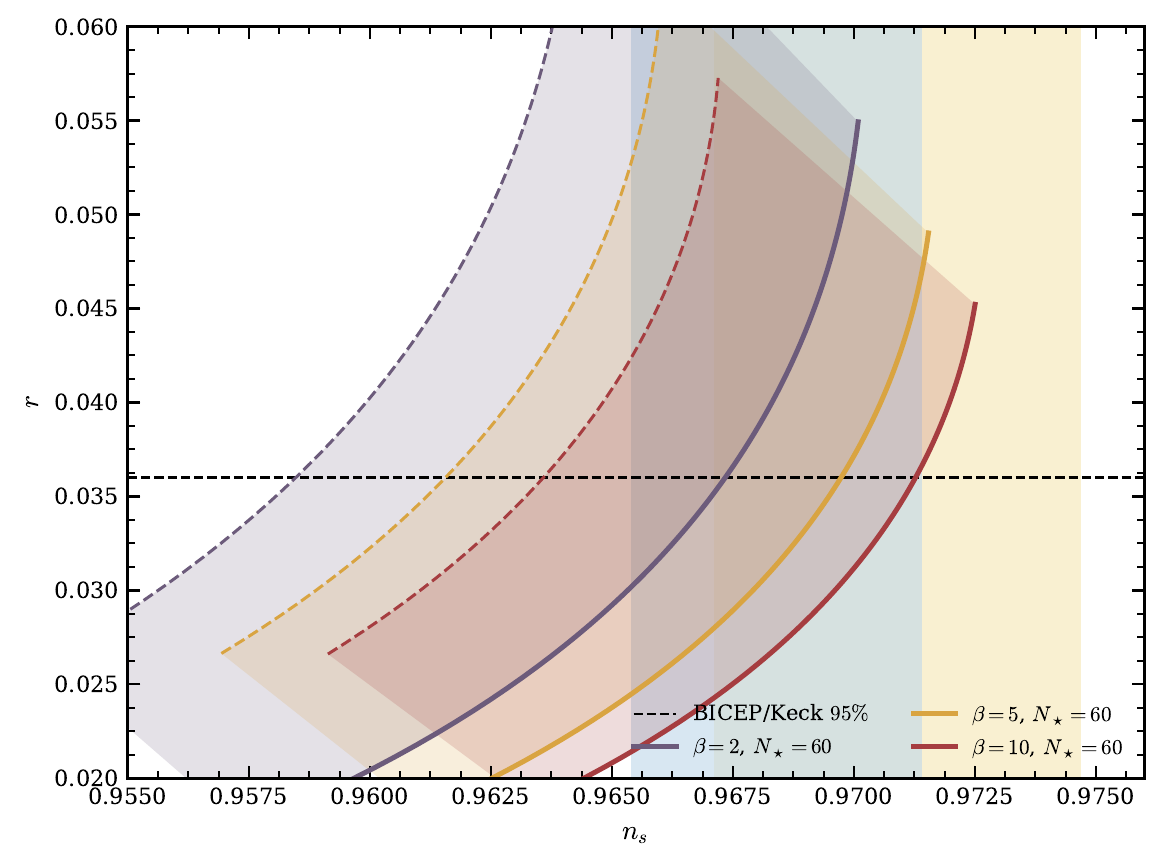}
    \caption{Robustness bands in the $(n_s,r)$ plane obtained by varying $N_\star$ between $50$ and $60$ for representative flattened branches. Solid curves show the $N_\star=60$ slice and dashed curves the $N_\star=50$ slice; the shaded area between them is the model band generated by the allowed $N_\star$ variation for each displayed $\beta$. The ACT+Planck and SPT-3G/CMB-SPA tilt intervals are shown as alternative $1\sigma$ anchors, and the horizontal dashed line is the BICEP/Keck bound.}
    \label{fig:nsrbands}
\end{figure}

For the numerical results we specialize to the tuned inflationary branch $B=0$. Two uses of the model should be distinguished. The fixed-$N_\star=60$ slices in Figs.~\ref{fig:nsr}--\ref{fig:nsrbands} remain useful as illustrative observational cross-sections of the closed-form EFT. The headline benchmark numbers quoted below, however, are evaluated on the  trough-reduced action of Eqs.~\eqref{eq:Gtr}--\eqref{eq:srtr} and are calibrated at $N_\star=56$ so as to lie on the positive-duration side of the constant-$\weff$ reheating map. The overall scale $\chi_0$ is fixed by the measured scalar amplitude,
\begin{equation}
A_s = \frac{U_\star}{24\pi^2\mpl^4\epsilon_\star},
\label{eq:As}
\end{equation}
with $A_s=2.101\times 10^{-9}$ at the pivot scale $k_\star=0.05\,\mathrm{Mpc}^{-1}$. Since the backreaction-resummed potential is closed-form, all model dependence is reduced to the pair $(\beta,f_a/\mpl)$ once $A_s$ is imposed, while the  trough metric induces only a mild numerical reshuffling of the constant-kinetic slices. This is consistent with the small size of Fig.~\ref{fig:metriccorr}, but the  trough reduction is nevertheless used for all benchmark tables and reheating quantities. The resulting presentation is therefore closer to a calibrated phenomenological benchmark than to a purely illustrative trajectory: the same EFT that fixes the flattening also fixes the theory-side correction associated with the curved trough itself.

Figure~\ref{fig:nsr} displays the model $n_s$--$r$ trajectories at fixed $N_\star=60$ for several values of $\beta$. The unflattened branch $\beta=0$ reproduces ordinary natural inflation. Increasing $\beta$ shifts the trajectories downward in $r$ at fixed $n_s$, as expected from the hilltop suppression factor in Eq.~\eqref{eq:hilltopfactor}. A useful quantitative consequence is that the BICEP/Keck bound can be translated into an illustrative fixed-tilt calibration of the same-sector flattening parameter. Solving the closed-form EFT at fixed $N_\star=60$ and fixed central tilt gives approximately
\begin{equation}
\begin{aligned}
\beta &\gtrsim 2.9 \quad \text{for}\quad n_s=0.9684, \\
\beta &\gtrsim 8.4 \quad \text{for}\quad n_s=0.9709,
\end{aligned}
\label{eq:thresholds}
\end{equation}
where the first number refers to the SPT-3G/CMB-SPA central value and the second to the ACT+Planck central value. Figure~\ref{fig:thresholdfig} shows these fixed-tilt calibrations directly. They are not likelihood bounds; they are consistency slices of the model on a fixed $N_\star=60$ trajectory.

A more conservative presentation, without turning the paper into a full MCMC study, is to widen the theory output from a single slice to a finite band. Figure~\ref{fig:nsrbands} therefore varies $N_\star$ over the interval $50$--$60$ for representative flattened branches and overlays the quoted ACT+Planck and SPT-3G/CMB-SPA $1\sigma$ tilt windows together with the BICEP/Keck tensor bound. The two observational anchors are treated as alternatives rather than as simultaneous constraints. Two features emerge. The downward shift of $r$ with increasing $\beta$ survives the full $N_\star$ variation, and once the observational input is presented as a finite $(n_s,r)$ window rather than as a single central value, the appropriate output of the theory is a robustness band rather than a one-number threshold. This is the appropriate interpretation of the analytic control achieved here: the paper provides prediction slices and band boundaries within the local EFT, not a full statistical fit to the CMB likelihoods~\cite{Planck:2018jri,Louis:2025fju,Camphuis:2025spt,BICEPKeck:2024txt}.

\begin{figure}[t]
    \includegraphics[width=\linewidth]{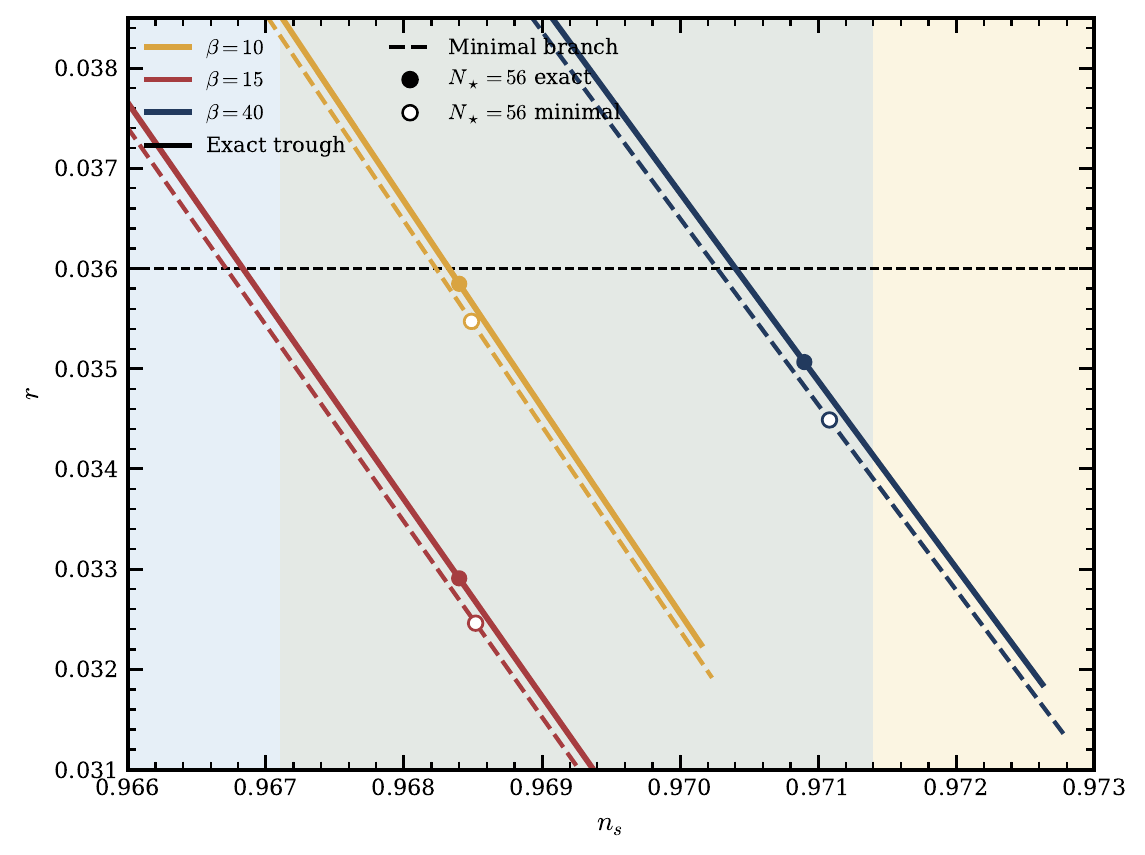}
    \caption{-trough (solid) and minimal constant-kinetic (dashed) predictions in the $(n_s,r)$ plane for the reheating-compatible benchmark branches. Filled markers indicate the exact $N_\star=56$ benchmark points and open markers the corresponding minimal-branch values. The shift is subdominant to the present observational widths but nonzero, providing a quantitative calibration of the reduced-theory uncertainty.}
    \label{fig:exactminimal}
\end{figure}

A further phenomenological gain of the  trough reduction is that it calibrates the residual theory-side ambiguity associated with the reduced one-field description itself. Figure~\ref{fig:exactminimal} overlays the -trough and minimal constant-kinetic predictions for the reheating-compatible benchmark branches over the observationally relevant interval $N_\star=50$--$60$. Across that window one finds
\begin{equation}
\begin{aligned}
|\Delta n_s|_{\rm max} &\lesssim 1.1\times10^{-4}\; (\beta=10), \\
                        &\lesssim 1.4\times10^{-4}\; (\beta=15), \\
                        &\lesssim 2.2\times10^{-4}\; (\beta=40), \\[3pt]
|\Delta r|_{\rm max}  &\lesssim 4.6\times10^{-4}\; (\beta=10), \\
                        &\lesssim 5.5\times10^{-4}\; (\beta=15), \\
                        &\lesssim 7.1\times10^{-4}\; (\beta=40).
\end{aligned}
\label{eq:obsbudget}
\end{equation}
while for the canonical reheating choice $\weff=0$ the corresponding constant-$\weff$ map shifts by only
\begin{equation}
\begin{aligned}
|\Delta N_{\rm re}|_{\rm max} &\lesssim 3.7\times10^{-2}\; (\beta=10), \\
                                &\lesssim 5.6\times10^{-2}\; (\beta=15), \\
                                &\lesssim 8.8\times10^{-2}\; (\beta=40).
\end{aligned}
\label{eq:rebudget}
\end{equation}
This observable-budget statement is useful beyond the present model: it identifies how much of the $(n_s,r,N_{\rm re})$ output is already fixed by the solvable local EFT and how much room remains for genuinely ultraviolet information to enter through microscopic trough deformations or explicit reheating couplings. In that sense the  trough action promotes the construction from a qualitative flattened potential to a quantitatively calibrated benchmark for confining, supersymmetric, or string-motivated embeddings~\cite{Planck:2018jri,Stein:2021yma,Green:2007gs,Cicoli:2010ha}.

Table~\ref{tab:bench} summarizes three reheating-compatible benchmark trajectories evaluated on the  trough-reduced action. Two are SPT-anchored branches with moderate and stronger flattening, while the third is an ACT-anchored branch with stronger backreaction chosen so that the benchmark remains inside the BICEP/Keck bound while keeping $N_{\rm re}>0$ in the constant-$\weff$ map. All three are calibrated at $N_\star=56$, slightly below their respective $N_{\rm re}=0$ boundaries. The resulting tensor amplitudes stay in the narrow interval $r\simeq0.033$--$0.036$, the running is consistently of order $-4.6\times10^{-4}$, and the adiabaticity diagnostics remain comfortably controlled. The  trough metric of Fig.~\ref{fig:metriccorr} shifts these benchmark observables at only the sub-percent to few-percent level relative to the minimal branch, but it is retained exactly throughout the benchmark analysis. The same-sector backreaction therefore continues to soften the tensor signal at fixed tilt while preserving explicit analytic control over the reduced one-field description.

\begin{table}[t]
\caption{Reheating-compatible  trough-reduced benchmark trajectories on the branch $B=0$, calibrated at $N_\star=56$. The last two columns give the minimum orthogonal-mode mass ratio and the maximum turn-rate ratio over the observable window $N=50$--$60$ for $b=2$; the corresponding trough-metric correction is displayed separately in Fig.~\ref{fig:metriccorr}.}
\label{tab:bench}
\begin{ruledtabular}
\tiny
\setlength{\tabcolsep}{2.5pt}
\begin{tabular}{lcccccc}
anchor & $\beta$ & $f_a/\mpl$ & $r$ & $\alpha_s$ & \shortstack{$\min(m_\perp^2/H^2)$\\$N=50$--$60$} & \shortstack{$\max(\Omega/H)$\\$N=50$--$60$} \\
\hline
SPT & 10 & 5.669 & 0.03585 & $-4.71\times10^{-4}$ & 7.86 & $6.43\times10^{-4}$ \\
SPT & 15 & 5.349 & 0.03291 & $-4.60\times10^{-4}$ & 7.08 & $7.12\times10^{-4}$ \\
ACT & 40 & 6.164 & 0.03507 & $-4.63\times10^{-4}$ & 6.05 & $7.53\times10^{-4}$ \\
\end{tabular}
\end{ruledtabular}
\end{table}

The benchmark excursions remain of the expected large-field size, $\Delta a/\mpl=f(\theta_\star-\theta_{\rm end})=8.0$--$9.6$, so the same-sector backreaction should be viewed as a dynamical flattening of natural inflation rather than as an independent ultraviolet resolution of field-range issues. The phenomenological value of the construction is different: once the local EFT is specified, the backreaction, the induced trough metric, the adiabaticity diagnostics, and the reheating map can all be tracked within one calculable framework. This makes the model useful as a benchmark template for ultraviolet completions, because a candidate completion must reproduce a correlated CMB-and-reheating pattern rather than only a single point in the $(n_s,r)$ plane~\cite{Planck:2018jri,Stein:2021yma,Nakayama:2008ip}.

\begin{figure}[t]
    \includegraphics[width=\linewidth]{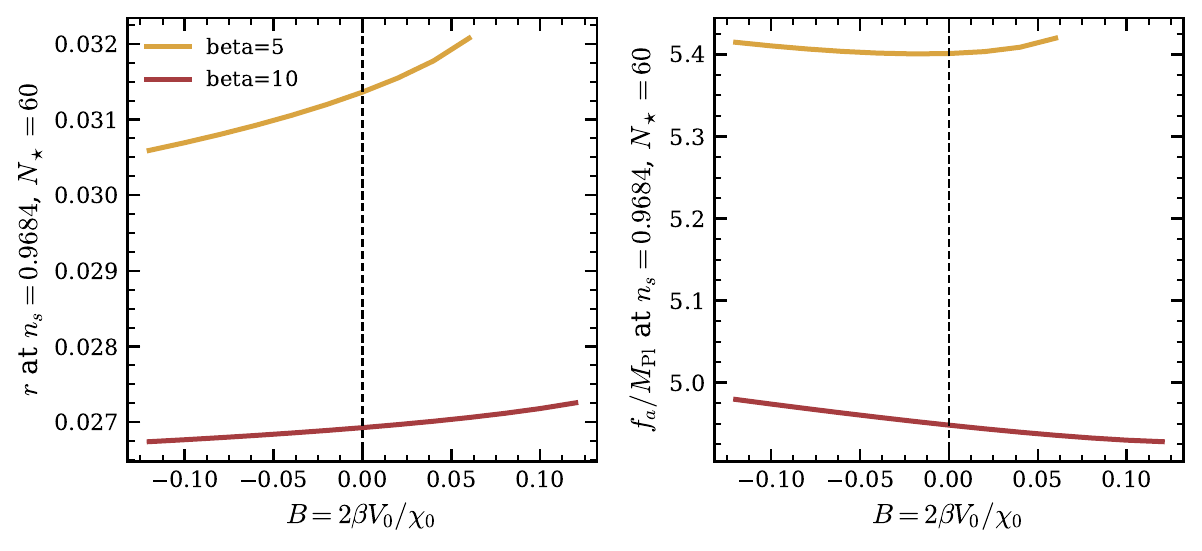}
    \caption{Sensitivity of the SPT-centered $N_\star=60$ slice to a modest nonzero vacuum offset $B=2\beta V_0/\chi_0$. For each displayed $B$, $f_a$ is retuned so that the backreacted potential reproduces $n_s=0.9684$. The resulting shifts in $r$ and $f_a/\mpl$ remain mild over the displayed interval, showing that the calibrated branch $B=0$ is not numerically singular.}
    \label{fig:Bsens}
\end{figure}

A second robustness issue is the role of the vacuum-offset parameter $B=2\beta V_0/\chi_0$. The full formulas in Sec.~\ref{sec:exact} keep $B$ explicit, while the benchmark figures above were drawn on the calibrated branch $B=0$. Figure~\ref{fig:Bsens} shows how the SPT-centered $N_\star=60$ slice moves when one retunes $f_a$ to keep $n_s=0.9684$ at small nonzero $B$. The result is mild over the displayed interval: for the $\beta=5$ branch and $-0.06\le B\le0.06$, the tensor amplitude shifts only from $r\simeq0.0309$ to $0.0321$ while $f_a/\mpl$ remains within about $0.4\%$ of its calibrated value; for the $\beta=10$ branch and $-0.10\le B\le0.10$, one finds $r\simeq0.0268$--$0.0272$ with $f_a/\mpl$ shifting by less than $1\%$. The $B=0$ branch should therefore be interpreted as a convenient calibration branch rather than as a numerically singular knife-edge, even though larger positive $B$ can eventually move a given fixed-tilt calibration outside the chosen benchmark window.

\section{Reheating relations for the backreacted potential}
\label{sec:reheating}

The reheating discussion can be stated without introducing any additional phenomenological couplings. We adopt the standard constant-$\weff$ treatment~\cite{Martin:2010kz,Dai:2014jja,Cook:2015vqa,Nakayama:2008ip}, but specialize all model-dependent quantities to the backreacted potential in Eq.~\eqref{eq:Ueff}. This approximation provides an averaged description of the post-inflationary expansion history and is standard in the reheating literature~\cite{Kofman:1994rk,Kofman:1997yn,Bassett:2005xm,Allahverdi:2010xz,Amin:2014eta,Lozanov:2019jxc}. In the present context it furnishes a reheating \emph{map} directly tied to the backreacted inflationary potential, while the underlying microphysical completion can be specified independently. Let $\rho_{\rm re}$ denote the energy density at the onset of radiation domination and $\rho_{\rm end}$ the density at the end of inflation. Energy conservation during reheating gives
\begin{equation}
\rho_{\rm re} = \rho_{\rm end}\,e^{-3(1+\weff)N_{\rm re}},
\label{eq:rhore}
\end{equation}
where $N_{\rm re}$ is the number of e-folds accumulated during reheating. Assuming thermalization at the end of reheating,
\begin{equation}
\rho_{\rm re}=\frac{\pi^2}{30}g_{\rm re}T_{\rm re}^4,
\label{eq:rhot}
\end{equation}
with $g_{\rm re}=106.75$ in the numerical examples. Entropy conservation between reheating and the present epoch implies
\begin{equation}
\frac{a_{\rm re}}{a_0} = \left(\frac{43}{11g_{s,\rm re}}\right)^{1/3}\frac{T_0}{T_{\rm re}},
\label{eq:entropy}
\end{equation}
where $g_{s,\rm re}=106.75$ is used below.

Combining Eqs.~\eqref{eq:rhore}--\eqref{eq:entropy} with the horizon-crossing condition $k_\star=a_\star H_\star$ gives
\begin{equation}
\ln\frac{k_\star}{a_0T_0} = -N_\star - N_{\rm re} + \frac13\ln\frac{43}{11g_{s,\rm re}} - \ln T_{\rm re} + \ln H_\star.
\label{eq:kmatch}
\end{equation}
Using Eq.~\eqref{eq:rhot} to eliminate $T_{\rm re}$ yields the constant-$\weff$ relation
\begin{equation}
\begin{aligned}
N_{\rm re} = \frac{4}{1-3\weff}\Bigg[&-N_\star - \ln\frac{k_\star}{a_0T_0}
+ \frac13\ln\frac{43}{11g_{s,\rm re}} \\
&- \frac14\ln\frac{30\rho_{\rm end}}{\pi^2 g_{\rm re}}
+ \ln H_\star\Bigg].
\end{aligned}
\label{eq:Nre}
\end{equation}
The reheating temperature then follows from Eqs.~\eqref{eq:rhore} and \eqref{eq:rhot}:
\begin{equation}
T_{\rm re} = \left(\frac{30\rho_{\rm end}}{\pi^2 g_{\rm re}}\right)^{1/4}
\exp\!\left[-\frac34(1+\weff)N_{\rm re}\right].
\label{eq:Tre}
\end{equation}
For the backreacted potential, the inflationary Hubble scale and the end-of-inflation energy density are fixed by
\begin{equation}
H_\star = \pi\mpl\sqrt{\frac{A_s r_\star}{2}},
\qquad
\rho_{\rm end}\simeq \frac32 U_{\rm end},
\label{eq:Hrhoend}
\end{equation}
where $U_{\rm end}$ is evaluated with Eq.~\eqref{eq:Ueff}. The relation $\rho_{\rm end}\simeq 3U_{\rm end}/2$ is the standard $\epsilon_H\simeq1$ estimate used in the reheating literature.

A particularly useful diagnostic is the boundary of instantaneous reheating, defined by $N_{\rm re}=0$. Setting Eq.~\eqref{eq:Nre} to zero gives
\begin{equation}
\begin{aligned}
N_\star^{(0)}=
&-\ln\frac{k_\star}{a_0T_0}
+\frac13\ln\frac{43}{11g_{s,\rm re}} \\
&-\frac14\ln\frac{30\rho_{\rm end}}{\pi^2 g_{\rm re}}
+\ln H_\star,
\end{aligned}
\label{eq:Nstar0}
\end{equation}
which is independent of $\weff$ because the boundary corresponds to vanishing reheating duration. Equation~\eqref{eq:Nstar0} provides a sharper and more predictive statement than quoting only a sign change in $N_{\rm re}$: once the inflationary branch is specified, it determines the maximal $N_\star$ compatible with conventional positive-duration reheating in the constant-$\weff$ map.

\begin{figure}[t]
    \includegraphics[width=\linewidth]{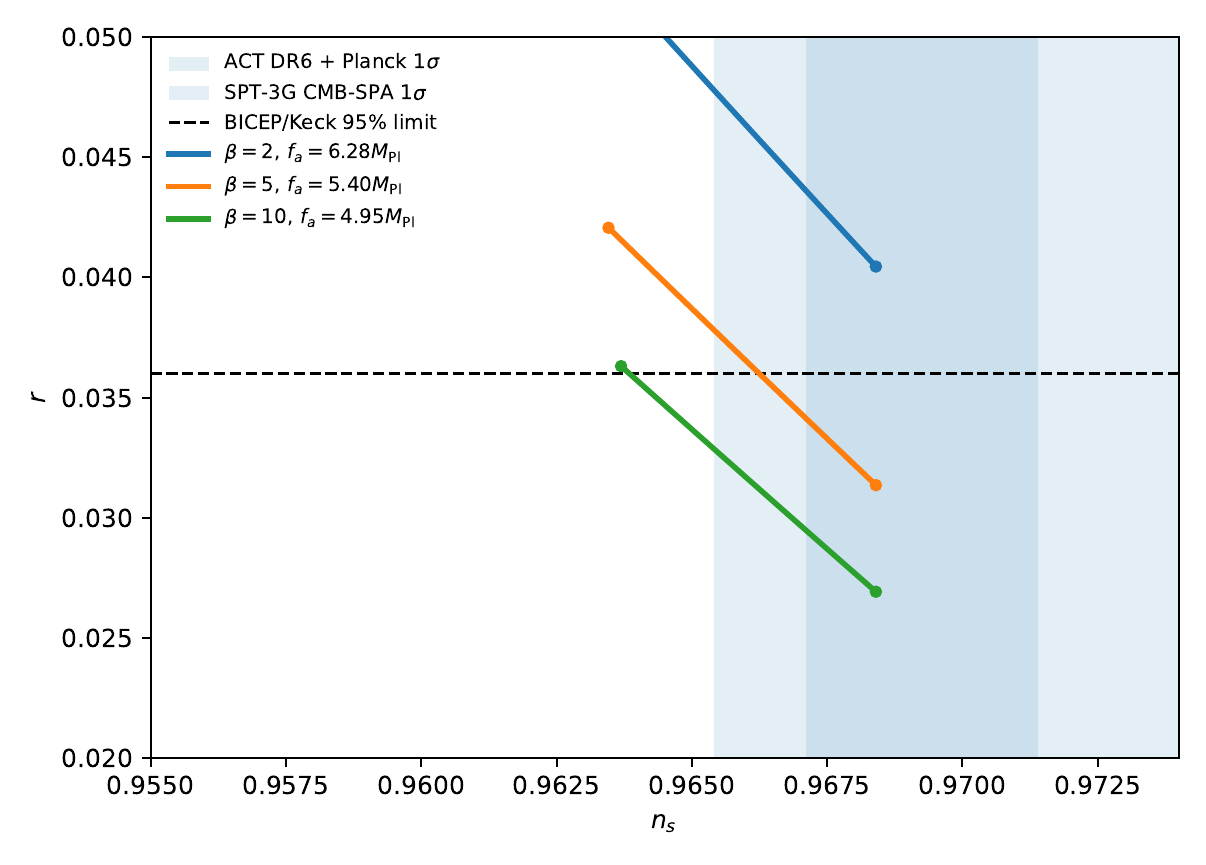}
    \caption{Model $n_s$--$r$ loci obtained by varying $N_\star$ from $50$ to $60$ for the reheating-compatible -trough benchmark branches. Filled markers indicate the adopted $N_\star=56$ benchmark points, while open markers indicate the corresponding $N_{\rm re}=0$ boundary points.}
    \label{fig:rehband}
\end{figure}

\begin{figure}[t]
    \includegraphics[width=\linewidth]{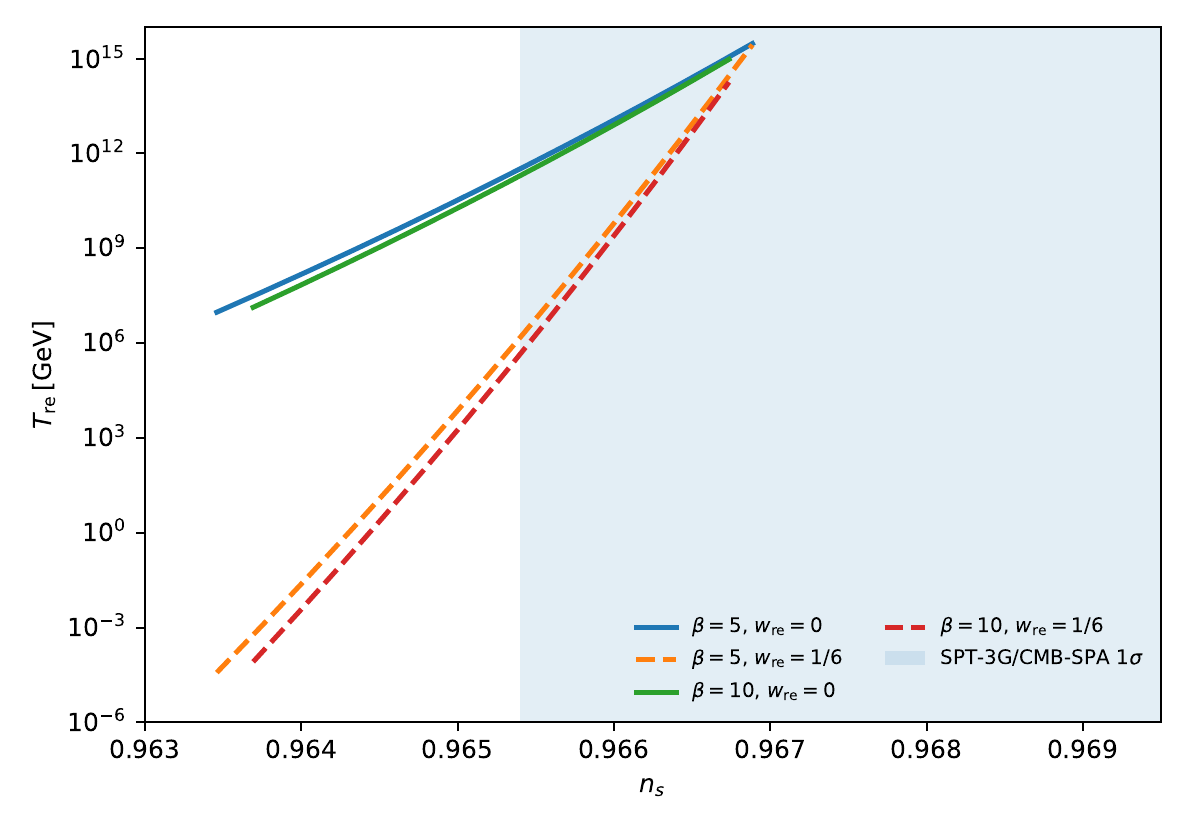}
    \caption{Reheating temperatures obtained from Eqs.~\eqref{eq:Nre} and \eqref{eq:Tre} for the reheating-compatible -trough benchmark branches and two representative mean equations of state, $\weff=0$ and $\weff=1/6$. Only the region with $N_{\rm re}\ge0$ is shown.}
    \label{fig:Trefig}
\end{figure}

The reheating consequences are shown in Figs.~\ref{fig:rehband} and \ref{fig:Trefig}. Figure~\ref{fig:rehband} presents the loci traced in the $n_s$--$r$ plane when $N_\star$ is varied between $50$ and $60$ for the reheating-compatible -trough benchmark branches; the filled markers indicate the adopted $N_\star=56$ benchmarks and the open markers show the corresponding $N_{\rm re}=0$ boundary points. Figure~\ref{fig:Trefig} gives the corresponding reheating temperatures after imposing $N_{\rm re}\ge0$. The main quantitative point is that the branches remain close to, but on the positive-duration side of, the conventional reheating boundary at the benchmark points actually used in the paper. For the $\beta=10$, $15$, and $40$ branches, Eq.~\eqref{eq:Nstar0} gives
\begin{equation}
\begin{aligned}
N_\star^{(0)}
&\simeq 56.49\; (\beta=10), \\
&\simeq 56.44\; (\beta=15), \\
&\simeq 56.42\; (\beta=40),
\end{aligned}
\label{eq:Nstar0num}
\end{equation}
which corresponds along the benchmark trajectories to
\begin{equation}
\begin{aligned}
 n_s^{(0)} &\simeq 0.96863,\quad r^{(0)}\simeq0.03537 \;\; (\beta=10), \\
 n_s^{(0)} &\simeq 0.96860,\quad r^{(0)}\simeq0.03252 \;\; (\beta=15),
\end{aligned}
\label{eq:Nrezeroa}
\end{equation}
for the SPT-anchored branches, and
\begin{equation}
 n_s^{(0)} \simeq 0.97110,\quad r^{(0)}\simeq0.03470 \;\; (\beta=40)
\label{eq:Nrezero}
\end{equation}
for the ACT-anchored branch. The benchmark points adopted in Table~\ref{tab:bench} sit slightly below these boundaries and therefore yield positive reheating durations already for the canonical choice $\weff=0$. In this sense the constant-$\weff$ post-processing no longer introduces a tension for the headline benchmark points; instead it identifies a narrow but nonzero reheating-compatible corridor adjacent to the observationally preferred slices. As usual, a dedicated microphysical reheating completion of the confining sector would refine this averaged map rather than replace the inflationary dynamics encoded in Eq.~\eqref{eq:Ueff}.

Representative reheating values are listed in Table~\ref{tab:reheating} directly at the adopted $N_\star=56$ benchmark points, for which all displayed branches already satisfy $N_{\rm re}>0$. The resulting temperatures are high, $T_{\rm re}\sim10^{14}$--$10^{15}$\,GeV for $\weff=0$, but they remain below the characteristic inflationary scale $\rho_{\rm end}^{1/4}$ and should be read as outputs of the averaged constant-$\weff$ map rather than as a replacement for a dedicated thermal history. Reheating scales of this order are not unusual when viable inflationary trajectories lie close to the short-duration or instantaneous-reheating boundary; values approaching $10^{15}$\,GeV are obtained in several otherwise consistent single-field and modified-gravity settings~\cite{Cook:2015vqa,Gialamas:2019lrv,Chien:2021nih}. At the same time, explicit ultraviolet completions can narrow the viable window: supersymmetric completions may be constrained by thermal or nonthermal gravitino production, while string-motivated embeddings can redistribute the inflaton energy into hidden sectors or moduli before the visible sector thermalizes~\cite{Asaka:2000zh,Kohri:2005wn,Kawasaki:2008qe,Hasegawa:2017sgg,Green:2007gs,Cicoli:2010ha,Kofman:2005yz,Brandenberger:2008aj,Barnaby:2009mc,Aldabergenov:2021ycx}. The temperatures quoted here should therefore be interpreted as EFT-consistent target scales for explicit reheating couplings rather than as universal UV predictions. To isolate the predictive core of the reheating map, Table~\ref{tab:reheatingcrit} separately lists the $N_{\rm re}=0$ boundary inferred from Eq.~\eqref{eq:Nstar0}.

\begin{table}[t]
\caption{Representative reheating solutions at the adopted -trough benchmark points $N_\star=56$. The quoted values follow directly from Eqs.~\eqref{eq:Nre} and \eqref{eq:Tre}.}
\label{tab:reheating}
\begin{ruledtabular}
\begin{tabular}{ccccc}
$\beta$ & $\weff$ & $n_s$ & $N_{\rm re}$ & $T_{\rm re}\,[\mathrm{GeV}]$ \\
\hline
10 & $0$   & 0.96840 & 1.986 & $7.51\times10^{14}$ \\
10 & $1/6$ & 0.96840 & 3.972 & $1.03\times10^{14}$ \\
15 & $0$   & 0.96840 & 1.754 & $9.08\times10^{14}$ \\
15 & $1/6$ & 0.96840 & 3.508 & $1.57\times10^{14}$ \\
40 & $0$   & 0.97090 & 1.708 & $9.81\times10^{14}$ \\
40 & $1/6$ & 0.97090 & 3.416 & $1.78\times10^{14}$ \\
\end{tabular}
\end{ruledtabular}
\end{table}

\begin{table}[t]
\caption{Boundary of instantaneous reheating, $N_{\rm re}=0$, for the -trough benchmark branches. The quoted values follow from Eq.~\eqref{eq:Nstar0} evaluated on the benchmark trajectories and are independent of $\weff$.}
\label{tab:reheatingcrit}
\begin{ruledtabular}
\begin{tabular}{cccc}
$\beta$ & $N_\star^{(0)}$ & $n_s^{(0)}$ & $r^{(0)}$ \\
\hline
10 & 56.494 & 0.96863 & 0.03537 \\
15 & 56.436 & 0.96860 & 0.03252 \\
40 & 56.425 & 0.97110 & 0.03470 \\
\end{tabular}
\end{ruledtabular}
\end{table}
\section{Discussion}
\label{sec:discussion}

The present construction has several features that are worth isolating at the level of interpretation and scope.

At the structural level, the backreacted same-sector potential in Eq.~\eqref{eq:Ueff} is structurally distinct from simply postulating a flattened single-field branch. The flattening is generated by the response of the heavy radial mode to the topological vacuum energy, not by the insertion of an external plateau operator. In this sense the construction is closer in spirit to heavy-field flattening~\cite{Dong:2010in} than to a purely phenomenological deformation of natural inflation. At the same time, because the topological vacuum energy and the heavy radial sector are attributed to the same confining dynamics, the mechanism also differs conceptually from string-motivated axion-monodromy flattening, where the backreacting heavy sector typically lives in compactification data rather than in a single four-dimensional confining EFT~\cite{McAllister:2008hb}.

The comparison with pure natural inflation is precise. In pure natural inflation, the flattening is encoded directly in the large-$N$ branch structure of the topological vacuum energy~\cite{Nomura:2017rik,Nomura:2017tm,Gatica:2026pni}. Here the flattening is dynamical and radial: the branch parameter $\beta$ measures the relative importance of the heavy-dilaton response. The resulting observables inhabit a similar region of the $n_s$--$r$ plane, but the control parameters and the local consistency conditions are different. The fixed-tilt calibrations in Eq.~\eqref{eq:thresholds}, together with the finite $N_\star$ robustness bands in Fig.~\ref{fig:nsrbands}, make this distinction sharper: the present paper supplies analytic prediction bands within the local EFT rather than a full likelihood-level comparison. The distinctive payoff is therefore one of control rather than of a unique smoking-gun observable: flattening, adiabaticity, and reheating consistency are tied together by explicit formulas inside one benchmark EFT.

For phenomenology, this matters because the benchmark is constrained by a correlated set of outputs rather than by the tensor amplitude alone. Once $A_s$ is imposed, the same local EFT fixes the amount of flattening, the induced trough metric, the adiabaticity diagnostics, and the location of the short-duration reheating boundary. This converts the model from a one-observable deformation of natural inflation into a calibrated target for ultraviolet constructions: a viable completion must reproduce a linked CMB-and-reheating pattern, not merely a single point in the $(n_s,r)$ plane~\cite{Planck:2018jri,Stein:2021yma,Nakayama:2008ip,Gialamas:2019lrv,Chien:2021nih}.

The treatment of the vacuum offset $V_0$ is explicit. Equations~\eqref{eq:etapiB} and \eqref{eq:mratioB} retain $V_0$ through the combination $B=2\beta V_0/\chi_0$, while Fig.~\ref{fig:Bsens} shows that small nonzero $B$ does not destroy the benchmark phenomenology. The displayed $B=0$ branch therefore serves as a convenient calibration slice for the inflationary vacuum energy, while Fig.~\ref{fig:Bsens} shows explicitly that nearby nonzero $B$ deformations preserve the qualitative phenomenology.

The same-sector flattening mechanism should be interpreted with the right degree of ambition. On the displayed branches it lowers $r$ at fixed $n_s$ and does so without adding an external plateau operator, but it does not by itself eliminate the need for a super-Planckian \emph{effective} axion range. The benchmark solutions retain $f_a/\mpl=\ord{5}$ and $\Delta a/\mpl=\ord{10}$, so the present construction is most naturally interpreted as a dynamically flattened realization of natural inflation within a controlled local EFT, with ultraviolet field-range questions delegated to the eventual microscopic completion~\cite{deLaFuente:2014aca}.

The scope of the effective theory should be stated with the same precision as its claims. The present analysis is carried out within a local two-field EFT with quadratic radial potential near the heavy minimum and exponential susceptibility, while the benchmark phenomenology is evaluated on the  trough-reduced action rather than on the minimal constant-kinetic truncation. This is closely aligned with the logic of curved-valley EFT reductions developed in the multi-field inflation literature~\cite{Shiu:2011qw,Cespedes:2012hu,Burgess:2012dz,Gong:2016uoa}. A full microscopic matching of the susceptibility to a specific confining completion, and a dedicated computation of postinflationary couplings to the Standard Model, remain natural next layers of model building rather than prerequisites for the local-EFT control problem isolated here. The control formulas in Eqs.~\eqref{eq:deltax}--\eqref{eq:defcond} make this statement more precise: they show how cubic/quartic radial corrections or modest departures from a pure exponential susceptibility feed into the trough and the effective potential without undoing the Lambert-$W$ backbone of the solution. Once the same-sector local EFT is specified, its inflationary dynamics, induced trough metric, adiabaticity diagnostics, and averaged reheating map can all be organized analytically around that benchmark.

\section{Conclusions}

We have presented a closed-form same-sector EFT realization of anomaly-inspired axion inflation based on a confining-sector model containing a heavy trace-anomaly mode and an axion coupled to the topological density. Eliminating the heavy radial mode resums the tree-level backreaction into a Lambert-$W$ potential for the axion. Within that local EFT, the solution yields a negative quartic harmonic, a suppressed hilltop curvature, an analytic orthogonal-mode mass ratio, and an analytic adiabatic-turn relation.

The phenomenological implications are correspondingly sharp. The fixed-$N_\star=60$ slices continue to show that same-sector backreaction lowers the tensor amplitude relative to ordinary natural inflation, while the  trough-reduced reheating-compatible benchmarks at $N_\star=56$ yield $r\simeq0.033$--$0.036$ with running $\alpha_s\simeq-(4.6$--$4.7)\times10^{-4}$ and positive reheating durations already for the canonical choice $\weff=0$. Figure~\ref{fig:nsrbands} shows that the downward shift of $r$ persists as a finite robustness band when $N_\star$ is widened to $50$--$60$ and the observational anchors are treated as separate $1\sigma$ windows rather than as simultaneous targets. Figure~\ref{fig:Bsens} further shows that modest nonzero $B$ shifts these benchmark slices only mildly, so the calibrated $B=0$ branch is not numerically singular. For the trough benchmark branches, the orthogonal mode remains parametrically heavy, the turn rate remains parametrically small, and the induced trough metric stays close to the minimal branch throughout the observable window.

The principal theoretical result is that a same-sector backreaction mechanism can render the flattening of a natural-inflation-like potential analytically tractable inside a local heavy-dilaton EFT. The resulting formulas are sufficiently rigid that present CMB data can already be translated into central-slice conditions on the backreaction parameter, lower bounds on the radial-response coefficient required for adiabatic control, and a sharp $N_{\rm re}=0$ reheating boundary for the benchmark branches. Equally important, the deformation formulas derived above provide a systematic bridge from the solvable benchmark EFT to nearby local completions without altering the central results established here. A microscopic confining derivation of the susceptibility and an explicit reheating completion define the natural next layer of refinement beyond this analytic benchmark.

\begin{acknowledgments}
I.K. acknowledges support from Zhejiang Normal University through a postdoctoral fellowship under Grant No.~YS304224924. TL is supported in part by the National Key Research and Development Program of China Grant No. 2020YFC2201504, by the Projects No. 11875062, No. 11947302, No. 12047503, and No. 12275333 supported by the National Natural Science Foundation of China, by the Key Research Program of the Chinese Academy of Sciences, Grant No. XDPB15, by the Scientific Instrument Developing Project of the Chinese Academy of Sciences, Grant No. YJKYYQ20190049, by the International Partnership Program of Chinese Academy of Sciences for Grand Challenges, Grant No. 112311KYSB20210012, and by the Henan Province Outstanding Foreign Scientist Studio Project, No.GZS2025008.

\end{acknowledgments}
\bibliography{refs}

\end{document}